\documentclass[sigconf,natbib=true,screen]{acmart}

\usepackage{graphicx}

\usepackage[utf8]{inputenc} 
\usepackage[T1]{fontenc}    
\usepackage{hyperref}       
\usepackage{url}            
\usepackage{booktabs}       
\usepackage{amsfonts}       
\usepackage{nicefrac}       
\usepackage{microtype}      
\usepackage{xcolor}         
\usepackage{multirow}
\usepackage{amsmath}        
\usepackage{todonotes}
\usepackage{enumerate}
\usepackage{enumitem}
\usepackage{cancel}
\usepackage{subcaption}

\usepackage{colortbl}      

\usepackage[ruled,linesnumbered]{algorithm2e}
\definecolor{highlight}{HTML}{C0C0C0}
\definecolor{highlight2}{HTML}{DCDCDC}
\definecolor{highlight3}{HTML}{F5F5F5}
\AtBeginDocument{%
  }

\acmSubmissionID{2705}

\copyrightyear{2024}
\acmYear{2024}
\setcopyright{acmlicensed}\acmConference[ASE '24]{39th IEEE/ACM International Conference on Automated Software Engineering }{October 27-November 1, 2024}{Sacramento, CA, USA}
\acmBooktitle{39th IEEE/ACM International Conference on Automated Software Engineering (ASE '24), October 27-November 1, 2024, Sacramento, CA, USA}
\acmDOI{10.1145/3691620.3695553}
\acmISBN{979-8-4007-1248-7/24/10}

\received{2024-06-07}
\received[Accepted]{2024-08-07}

\begin{document}

\title{A Joint Learning Model with Variational Interaction for Multilingual Program Translation}


\author{Yali Du}
\affiliation{%
    \department{National Key Laboratory for Novel Software Technology}
    \department{School of Artificial Intelligence}
    \institution{Nanjing University}
    \country{}
}
\email{duyl@lamda.nju.edu.cn}

\author{Hui Sun}
\affiliation{%
    \department{National Key Laboratory for Novel Software Technology}
    \department{School of Artificial Intelligence}
    \institution{Nanjing University}
    \country{}
}
\email{sunh@lamda.nju.edu.cn}

\author{Ming Li}
\authornote{Ming Li is the corresponding author.}
    \affiliation{%
    \department{National Key Laboratory for Novel Software Technology}
    \department{School of Artificial Intelligence}
    \institution{Nanjing University}
    \country{}
}
\email{lim@lamda.nju.edu.cn}


\begin{abstract}
Programs implemented in various programming languages form the foundation of software applications.
To alleviate the burden of program migration and facilitate the development of software systems, automated program translation across languages has garnered significant attention.
Previous approaches primarily focus on pairwise translation paradigms, learning translation between pairs of languages using bilingual parallel data. 
However, parallel data is difficult to collect for some language pairs,
and the distribution of program semantics across languages can shift,
posing challenges for pairwise program translation.
In this paper, we argue that jointly learning a unified model to translate code across multiple programming languages is superior to separately learning from bilingual parallel data. 
We propose Variational Interaction for Multilingual Program Translation~(VIM-PT), a disentanglement-based generative approach that jointly trains a unified model for multilingual program translation across multiple languages. 
VIM-PT disentangles code into language-shared and language-specific features, using variational inference and interaction information with a novel lower bound, then achieves program translation through conditional generation.
VIM-PT demonstrates four advantages: 
1) captures language-shared information more accurately from various implementations and improves the quality of multilingual program translation,
2) mines and leverages the capability of non-parallel data,
3) addresses the distribution shift of program semantics across languages, 
4) and serves as a unified model, reducing deployment complexity.

\end{abstract}

%
%



\maketitle

\section{Introduction}
Programs form the foundation of computer applications. 
Various programming languages have been invented to address diverse requirements.
Developing a new application from scratch is time- and labor-intensive,
whereas referring to and combining existing code is more efficient. 
Unfortunately, when combining programs written in different programming languages, the developer usually suffers from the intensive labor of manually translating programs from one programming language to another. For example, many industries spend hundreds of millions of dollars to convert code written in older programming languages~(e.g., FORTRAN and COBOL) to newer ones~(e.g., Java, C++)~\cite{Transcoder}. To alleviate the burden of program migration and facilitate the development of software systems, program translation, which aims to translate the program from one programming language to another automatically, has drawn significant attention in the software mining community~\cite{codetran5,codetran6,codetran4,codetran3,codetran2,codetran1,Transcoder,must,pt,pt2}.

The booming development of machine learning coupled with the availability of an extensive parallel corpus of programs has led to a remarkable enhancement in the performance of program translation.
For example, \citet{NguyenNN13, NguyenNN15} attempted to translate code across languages, leveraging phrase-based statistical machine translation and grammatical rules.
Recently, due to the potent power of representation learning in the deep neural networks,
most modern program translation approaches based on deep neural networks, such as sequence-to-sequence models, have advanced the state-of-the-art performance to a new level~\cite{tree2tree, GraphCodeBERT, pscpt}.

\begin{figure}[t]
 \vspace{-0.1cm}
    \centering
    \includegraphics[width=0.9\linewidth]{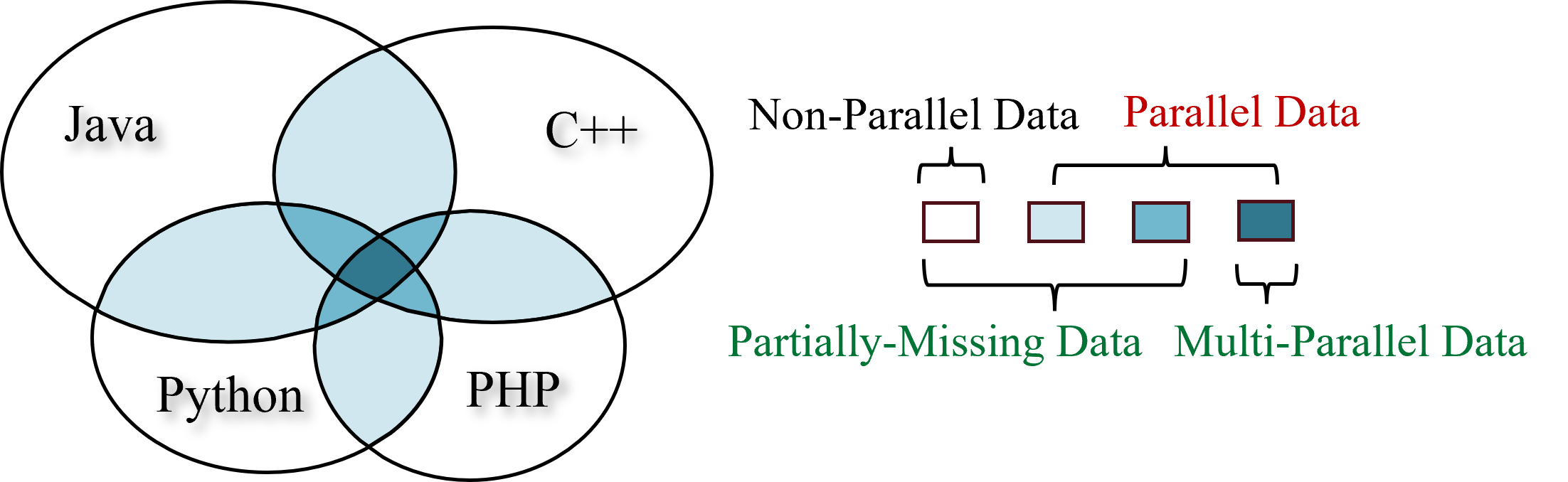}
    \caption{An example of a multilingual dataset.}
    \label{fig:dataset_example}
    \vspace{-0.6cm}
\end{figure}

Nevertheless, most existing approaches primarily focus on pairwise translation, lacking exploration in multilingual program translation.
Although pairwise approaches can be employed for each language pair separately, these approaches cannot leverage knowledge from languages outside the current pair,
posing a challenge for learning across a pair of languages with only a few parallel data.
As shown in Fig.~\ref{fig:dataset_example}, 
previous approaches, including the multilingual program translation approach by~\citet{must}, are trained only with bilingual parallel data.
However, these parallel data are sparse in practical datasets. 
For instance, the multilingual dataset \textit{CoST}~\cite{must} contains only 12.56\% bilingual parallel data among all possible pairs. In addtion, as shown in Fig.~\ref{fig:cost_pro_pairs} and Fig.~\ref{fig:cost_sni_pairs}. the distribution of different language pairs is imbalanced.
%

In this paper, we argue that jointly training a unified model to translate code across multiple programming languages is superior to separately training with pairwise data.
A program implemented in different programming languages should exhibit the invariant language-shared semantics, i.e., perform the same tasks.
Thus, joint training on multiple different implementations of a program can complement and enhance the learning of the language-shared representation.
By constructing the unified language-shared latent space, the rich resource languages can benefit the low resource languages in the translation.
However, joint learning involves a new challenge: as shown in Fig.~\ref{fig:dataset_example}, practical datasets are always semi-parallel, which contain only a few complete multi-parallel samples that include all languages, while most are partially missing implementations in some programming languages.
To tackle the issue of semi-parallel data, utilizing semi-supervised learning techniques is intuitive. 
Generative semi-supervised learning offers a solution: all data, including partially missing and multi-parallel data, can be generated from a latent distribution.
All these data can be used to learn the latent distribution and generative process with an Expectation–Maximization~(EM)-based algorithm~\cite{kingma2014semi}.
Once the latent distribution and generative process are learned, program translation can be achieved with a conditional generative process.

Building upon the above idea, we propose an innovative approach called Variational Interaction for Multilingual Program Translation~(VIM-PT), a disentanglement-based generative approach.
VIM-PT disentangles each code into \textbf{language-shared} features and \textbf{language-specific} features based on information theory.
The language-shared features should be task-specific and language-invariant like the functional semantics of the program, while the language-specific features should be 
language-specific and task-invariant like the grammar or syntax of the programming language.
Therefore, each code can be generated through the interplay of these features. 
Specifically, we employ variational inference to learn the generative process, where the prior distributions of both features follow a normal Gaussian distribution.
To enforce disentanglement, there will be a new variational interaction bound for the objective, 
compared to the Evidence Lower Bound~(ELBO) in traditional variational inference, which includes three additional terms from interaction information.
By optimizing this new variational interaction bound, VIM-PT learns to generate code based on disentangled features. 
This enables program translation from a source language to a target language by using the language-shared feature from the source language and the language-specific feature sampled from the prior distribution of the target language.

Compared to previous approaches, VIM-PT demonstrates four strengths:
1) More precisely and completely captures the language-shared feature by jointly learning from the various views of multiple implementations and improves the quality of multilingual program translation.
2) Effectively leverages the power of non-parallel and partially missing data using a generative framework.
3) Addresses the distribution shift of semantics across multiple languages. Different languages are often used for different tasks, such as Python for data science and machine learning, and JavaScript for front-end development, resulting in a distribution shift of semantics in collected data. 
VIM-PT addresses this by using a conditional generative framework to complete missing implementations in some languages.
4) VIM-PT is a unified model across different translation pairs, reducing the total parameters and making deployment more convenient, especially on edge devices.
To evaluate the effectiveness of the VIM-PT, an extensive experiment is conducted on the widely used dataset, which includes 7 general programming languages~(i.e., C, C\#, C++, Java, Javascript, PHP, and Python). It can be observed that VIM-PT performs performance gains widely over the state-of-the-art approaches.

We highlight our contributions in three key aspects:
\begin{enumerate}
    \item 
        We argue that jointly training a unified model to translate code across multiple programming languages is superior to training with pairwise data separately.
    \item 
        We propose VIM-PT, a disentanglement-based generative approach, for joint learning with variational interaction in multilingual program translation.
    \item 
        We conduct extensive experiments to demonstrate the effectiveness of VIM-PT and the necessity of jointly learning a unified model in multilingual program translation.
\end{enumerate}

\section{Related Works}

\subsection{Program Translation}
To meet some specific requirements, rule-based translation methods have been developed~\cite{rule1,rule2,rule3,rule4}.
Further, some works applied phrase-based statistical machine translation techniques to program translation~\cite{KaraivanovRV14,NguyenNN15,NguyenNN13,NguyenNN16a,AllamanisBDS18,pt6,swt,oda2015learning}, which leveraged grammatical structures of programming languages for code migration.  
Recently, various deep learning techniques were employed to program translations~\cite{LampleCDR18,babel,pt5}, which can be distinguished into pairwise program translation methods~\cite{tree2tree, GraphCodeBERT, pscpt,yang2024exploring} and multilingual program translation methods~\cite{must} by using bilingual or multilingual corpus in the supervised learning.

The pairwise approaches trained one model for each translation direction independently~\cite{GraphCodeBERT,pairtran1,pairtran2}.
However, a problem encountered by many existing models is that program translation datasets are usually not balanced in size for all the languages. Some languages may have much less parallel data than others. Less parallel training data can significantly affect the translation performance of low-resource languages. Therefore, some works leveraged multilingual training to improve the performance of low-resource languages, which is defined as multilingual program translation.
Aligning the shared semantics of multi-parallel data is the key to utilizing the rich resource to benefit the low resource. However, the existing multilingual program translation model MuST-PT~\cite{must} is limited to the pairwise training process and does not take full advantage of the multi-parallel data and partially missing data (e.g., non-parallel data).

\subsection{Exploring Non-Parallel Data for Translation}
The quality of translation systems highly depends on the quality of the available parallel data. However, for most languages, parallel resources are rare or nonexistent. 
Some approaches have investigated the use of non-parallel data to improve existing program translation systems~\cite{Transcoder,transcoder2,transcoder3,DOBF,codebert,codet5}.
For instance, \citet{DOBF} proposed DOBF that leverages the structural aspect of coding languages and pre-trains a model to recover the original version of obfuscated source code. 
\citet{codebert} proposed CodeBERT, which is a Bert-like model pre-trained in an open-source GitHub repository and used in many down-stream tasks, and then much relevant research emerged~\cite{GraphCodeBERT,semanticcodebert,sgattention}. 
\citet{Transcoder} proposed TransCoder, which trained a translation model only using the monolingual corpus with back-translation and denoising auto-encoding.

A common shortcoming of the above methods is the language-shared and language-specific representations among different languages are not disentangled in the learning. As shown in many researches, the disentanglement is crucial in such a way that every factor of variation is captured in the right part of the representation so that the partially missing data can be better utilized~\cite{mirror,hwang2020variational,su2018variational,eikema2018auto,pagnoni2018conditional,li2009variational,SetiawanSNP20,sheng2020introvnmt}.
For example, in neural machine translation, \citet{mirror} proposed the mirror-generative approach, which is a single unified architecture in which both translation models and language models share the same latent semantic space, therefore both translation directions can learn from non-parallel data. 
Yet the method is limited to bilingual translation and requires extra language models with more space overhead~\cite{mirror}. 
Inspired by the discussion on cross-domain disentanglement~\cite{hwang2020variational}, we expand the variational disentanglement to the multi-domain, and it is the first attempt to explore the variational disentanglement in multi-lingual translation among programming languages.

\section{Method}

The method is discussed in this section, including the generative model by variational inference, enforcing disentanglement by interaction information, the overall framework of VIM-PT, and training the joint model from both multi-parallel samples and partially missing samples.

\subsection{Generative Model via Variational Inference} 
Consider a sample program implemented in \(N\) languages, denoted as \((x^1, x^2, \cdots, x^N) \sim p_D(x^1, x^2, \cdots, x^N)\), where \(x^i\) represents the code written in the \(i\)-th programming language.
All of these \(N\) codes exhibit the same semantics, i.e., perform the same task,
while each code \(x_i\) involves the language-specific grammatical style.
From the perspective of disentanglement, each code is generated through the interplay of a language-shared feature, denoted as \(z^s\in Z^s\), and a language-specific feature, denoted as \(z^i \in Z^i\) for the \(i\)-th programming language.

\begin{figure}[!h]
    \centering
    \includegraphics[width=0.85\linewidth]{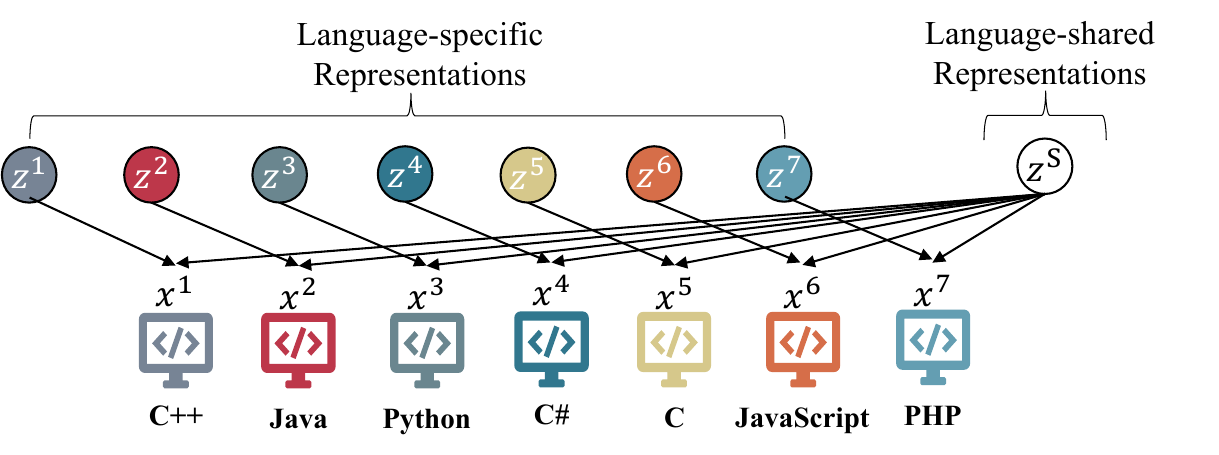}
    \caption{\label{fig:generation}
        Generative Process.
    }
\end{figure}
As illustrated in Fig.~\ref{fig:generation}, the generative process involves two steps:
1) The shared feature \(z^s\) is generated from the prior distribution \(p_{\theta^*}(z^s)\), while the language-specific feature \(z^i\) is generated from \(p_{\theta^*}(z^i)\).
2) Subsequently, the code \(x^i\) is generated from the conditional distribution \(p_{\theta^*}(x^i|z^s, z^i)\).
Following~\citet{vae}, these prior distributions \(p_{\theta^*}(z^s)\) and \(p_{\theta^*}(z^i)\), as well as the conditional distribution \(p_{\theta^*}(x^i|z^s, z^i)\), are drawn from parametric families of distributions \(p_{\theta}(z^s)\), \(p_{\theta}(z^i)\), and \(p_{\theta}(x^i|z^s, z^i)\), respectively.
Therefore, as per~\citet{vae}, the generative process can be formally expressed through the marginal likelihood of the joint distribution in the multilingual Bayesian network:
\begin{equation}\label{eq:gen_likelihood}
    p_{\theta}(x^1,x^2,\cdots,x^N) =\hspace{-0.5mm} \int \hspace{-0.5mm} p_\theta (z^s) dz^s  \prod_{i=1}^{N} p_\theta (z^i) p_\theta (x^i|z^i,z^s) dz^i \,.
\end{equation}

\begin{figure}[!h]
    \centering
    \includegraphics[width=0.85\linewidth]{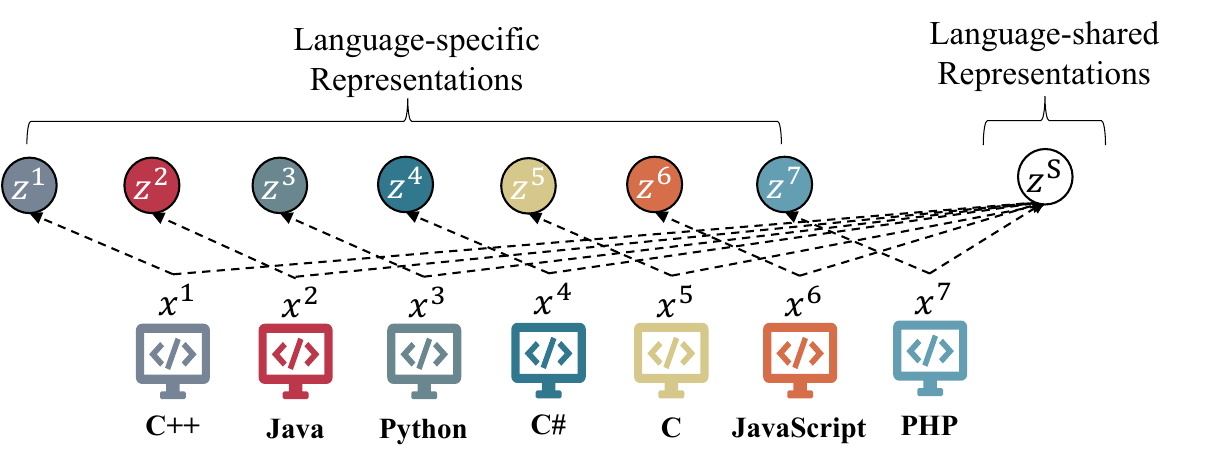}
    \caption{\label{fig:inference}
        Inference Process.
    }
\end{figure}

While Eq.~\ref{eq:gen_likelihood} is intractable, variational inference based on Variational Autoencoders (VAE) is typically employed~\cite{vae, kingma2014semi, hwang2020variational}. 
Considering the inference process in variational inference, as shown in Fig.~\ref{fig:inference}, the language-shared feature \(z^s\) can be jointly inferred from all codes implemented in various programming languages, while the language-specific grammatical style feature \(z^i\) is inferred from the code implemented in the \(i\)-th programming language.

Thus, the inference process can be formally expressed via the posterior distribution \(p_\theta (z^1, z^2, \cdots, z^N, z^s | x^1, x^2, \cdots, x^N)\), which can be approximated by:
\begin{equation}\begin{aligned}\label{eq:inf_posterior}
    & q_\phi(z^1, z^2, \cdots, z^N, z^s | x^1, x^2, \cdots, x^N) = \\
    & \quad\ q_{\phi}(z^1|x^1) q_{\phi}(z^2|x^2) \cdots q_{\phi}(z^N|x^N) q_{\phi}(z^s|x^1, x^2, \cdots, x^N) \,.   
\end{aligned}\end{equation}

By combining Eq.~\ref{eq:gen_likelihood} and Eq.~\ref{eq:inf_posterior}, and omitting subscripts $\theta$ and $\phi$ for brevity, the log-likelihood of the joint distribution can be lower-bounded using the ELBO as:
\begin{align}\label{eq:ELBO} 
        & \log p(x^1, x^2, \cdots, x^N)  \notag \\
        & \quad \geq \mathbb{E}_{q(z^1, \cdots, z^N, z^s | x^1, \cdots, x^N)}\left[\log \frac{p(x^1, \cdots, x^N, z^1, \cdots, z^N, z^s)}{q(z^1, \cdots, z^N, z^s | x^1, \cdots, x^N)}\right] \notag \\
        & \quad = \sum_{i=1}^N \mathbb{E}_{q(z^i | x^i) q(z^s | x^1, \cdots, x^N)} \left[\log p(x^i | z^i, z^s)\right] \\
        & \quad\  - \sum_{i=1}^N D_{KL}\left[q(z^i | x^i) \| p(z^i)\right] 
            - D_{KL}\left[q(z^s | x^1, \cdots, x^N) \| p(z^s)\right] \,, \notag 
\end{align}
where \(D_{KL}\) represents the KL-divergence.
Specifically, in variational inference, the prior distribution \(p(z^\cdot)\) is typically a Gaussian distribution, such that \(p(z^\cdot)\sim\mathcal{N}(\mathbf{0},\mathbf{I})\), while the approximate posterior \(q(z|x)\) can be reparameterized as \(q(z|x)\sim\mathcal{N}(\mu(z|x),\sigma(z|x))\).
By relying on a reparameterization trick~\cite{vae}, we can now jointly train all the components using gradient-based algorithms.

We formulate the generative process in Fig.~\ref{fig:generation} using Eq.~\ref{eq:gen_likelihood} and the inference process in Fig.~\ref{fig:inference} using Eq.~\ref{eq:inf_posterior}, then combine them to bound the log-likelihood with the ELBO in Eq.~\ref{eq:ELBO}. 
However, maximizing the ELBO in Eq.~\ref{eq:ELBO} does not control the disagreement in inference.
Disentanglement will be enforced using interaction information from information theory, as described by~\citet{hwang2020variational}, which will be discussed in detail next.

\subsection{Disentanglement~by~Interaction~Information}

In ideal disentanglement, the decomposed features should satisfy two key properties:
1) The language-shared feature \(z^s \in Z^s\) and the language-specific grammatical style feature \(z^i \in Z^i\) should avoid capturing redundant information.
2) For any code implemented in the \(i\)-th language, \(x^i \in X^i\), the common information between \(x^i\) and others (\(x^1, \cdots, x^{i-1}, x^{i+1}, \cdots, x^N\)) should be the language-shared information, i.e., \(z^s \in Z^s\).

To achieve the first property, we employ the mutual information \(I(\cdot; \cdot)\) between \(Z^s\) and \(Z^i\) to quantify redundancy:
\begin{equation}\label{eq:dis_p1}
    I(Z^i; Z^s) = -I(X^i; Z^i, Z^s) + I(X^i; Z^i) + I(X^i; Z^s) \,.
\end{equation}
The proof of this equation can be found in Appendix~\ref{IZIZS}.

To achieve the second property, we use interaction information, a generalization of mutual information among three or more random variables, to quantify the common information among the code implemented in the \(i\)-th programming language, the codes implemented in other languages, and the language-shared features. 
For brevity, we denote \(\{x^1, \cdots, x^{i-1}, x^{i+1}, \cdots, x^N\}\) as \(\overline{x_i}\) and \(\{X^1, \cdots, X^{i-1}, X^{i+1}, \cdots, X^N\}\) as \(\overline{X_i}\).
Formally, the common information can be expressed as:
\begin{equation}\label{eq:dis_p2}
    I(X^i; \overline{X_i}; Z^s) = \ I(X^i; Z^s) - I(X^i; Z^s | \overline{X_i}) \,,
\end{equation}
which can be derived from the definition of the interaction information between three random variables.

Then, we combine Eq.~\ref{eq:dis_p1} and Eq.~\ref{eq:dis_p2} to enforce disentanglement:
\begin{equation}\label{eq:disent} 
    I(X^i; \overline{X_i}; Z^s) - I(Z^i; Z^s) = -I(X^i;Z^s|\overline{X^i}) + I(X^i; Z^i,Z^s) - I(X^i;Z^i) \,.
\end{equation}

Unfortunately, directly maximizing Eq.~\ref{eq:disent} is still intractable.
Let us analyze it term by term.

\begin{itemize}[leftmargin=*]
\item The first term involves \( q(z^s|\overline{x^i}) = \int p_D(x^i|\overline{x^i})q(z^s|x^1,\cdots,x^N)dx^i\), where \(p_D(x^i|\overline{x^i})\) is unknown.
Therefore, using the variational distribution \(r^i(z^{s}|x^i)\), the first term can be lower bounded as follows:
\begin{align}
    & -I(X^i;Z^s|\overline{X^i}) = -\mathbb{E}_{p_D(x^1,\cdots, x^N)q(z^s|x^1,\cdots,x^N)}\left[
            \log \frac{q(z^s|x^1,\cdots,x^n)}{q(z^s|\overline{x^i})}
        \right] \notag \\
    & \quad = -\mathbb{E}_{p_D(x^1,\cdots, x^N)q(z^s|x^1,\cdots,x^N)} \left[
            \log \frac{q(z^s|x^1,\cdots,x^N) r^i(z^{s}|x^i)}{q(z^s|\overline{x^i}) r^i(z^s|x^i)}
        \right] \notag \\
    & \quad = -\mathbb{E}_{p_D(x^1,\cdots, x^N)}\left[D_{KL}\left[
            q(z^s|x^1,\cdots,x^N) \| r^i(z^s|x^i)
        \right]\right] \label{eq:disent_1st} \\
    & \quad \quad  + \mathbb{E}_{p_D(\overline{x^i})}\left[D_{KL}\left[
            q(z^s|x^i) \| r^i(z^s|x^i)
        \right]\right] \notag \\
    & \quad \geq -\mathbb{E}_{p_D(x^1,\cdots, x^N)}\left[D_{KL}\left[
            q(z^s|x^1,\cdots,x^N) \| r^i(z^s|x^i)
        \right]\right] \,. \notag
\end{align} 
Thus, when maximizing $-I(X^i;Z^s|\overline{X^i})$ using Eq.~\ref{eq:disent_1st}, the variational distribution $r^i(z^s|x^i)$ will be learned to fit $q(z^s|x^1,\cdots,x^N)$.

\item Meanwhile, the second term \(I(X^i; Z^i,Z^s)\) involves \(q(x^i|z^i,z^s) = \frac{q(z^i,z^s|x^i)p_D(x^i)}{\int p_D(x^i,\cdots,x^N) q(z^i,z^s|x^i,\cdots,x^N) dx^i,\cdots, dx^N}\), \ where \(p_D(x^1,\cdots,x^N)\) and \(p_D(x^i)\) are unknown.
However, it can be lower-bounded using the generative distribution \(p(x^i|z^i,z^s)\) as follows:
\begin{align}\label{eq:disent_2nd}
    & I(X^i; Z^i,Z^s) = \mathbb{E}_{q(z^i,z^s|x^i) p_D(x^i)} \left[ \log \frac{q(x^i|z^i,z^s)}{p_D(x^i)} \right] \notag \\
    & \quad = H(X^i) + \mathbb{E}_{q(z^i,z^s|x^i) p_D(x^i)} \left[ \log p(x^i|z^i,z^s) \right] \notag \\
    & \quad \quad + \mathbb{E}_{q(z^i,z^s)}\left[ D_{KL} \left[ q(x^i|z^i,z^s) \| p(x^i|z^i,z^s) \right]\right] \\
    & \quad \geq H(X^i) + \mathbb{E}_{q(z^i,z^s|x^i)p_D(x^i)} \left[ \log p(x^i|z^i,z^s) \right] \notag \\
    & \quad =  H(X^i) + \mathbb{E}_{p_D(x^i,\cdots,x^N)q(z^i|x^i)q(z^s|x^i,\cdots,x^N)} \left[ \log p(x^i|z^i,z^s) \right] \notag
\end{align}
where \(H(\cdot)\) represents the Shannon entropy.

\item The third term \(-I(X^i;Z^i)\) involves \(q(z^i) = \int p_\mathcal{D}(x^i)q(z^s|x^i)dx\), where \(p_\mathcal{D}(x^1)\) is unknown. We employ the Variational Information Bottleneck (VIB)~\cite{vib} to lower bound it:
\begin{equation}\begin{aligned}
    - I(X^i;Z^i) & = -\mathbb{E}_{p_D(x^i)}\left[D_{KL}\left[q(z^i|x^i) \| q(z^i)\right]\right] \\
    & \geq -\mathbb{E}_{p_D(x^i)}\left[D_{KL}\left[q(z^i|x^i) \| p(z^i)\right]\right]
\end{aligned}\end{equation} 
Here we use $-\mathbb{E}_{p_D(x^i)}\left[D_{KL}\left[q(z^i|x^i) \| p(z^i)\right]\right]$ as its lower bound with the generative distribution $p(z^i)$ defined as the standard Gaussian.

\end{itemize}

\begin{figure*}[t]
    \centering
    \includegraphics[width=0.96\linewidth]{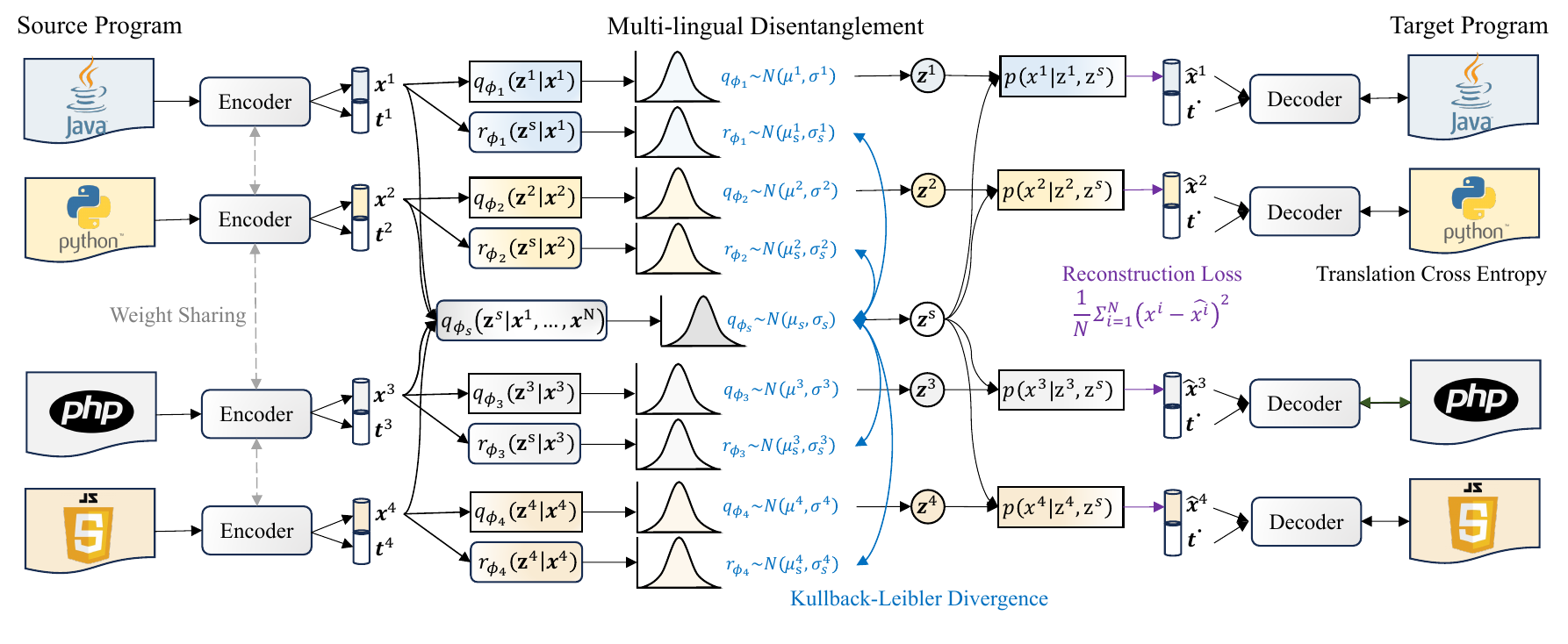}
    \caption{The overall framework.}
    \label{framework}
\end{figure*}

Subsequently, we can derive a lower bound for Eq.~\ref{eq:disent} by combining the derived equations. Therefore, we can enforce disentanglement across multiple programming languages by maximizing:
\begin{align}\label{eq:disent_bound}
    & \sum^N_{i=1}\left[ I(X^i;\overline{X^i};Z^s) - I(Z^i;Z^s) \right] \notag \\
    & \quad = \sum^N_{i=1} \left[ -I(X^i;Z^s|\overline{X^i}) + I(X^i; Z^i,Z^s) - I(X^i;Z^i) \right] \notag \\
    & \quad \geq \mathbb{E}_{p_D(x^1,\cdots,x^N)} \sum_{i=1}^N \mathbb{E}_{q(z^i|x^i)q(z^s|x^1,\cdots,x^N)} \left[ 
            \log p(x^i|z^i,z^s)
        \right] \notag \\
    & \quad\quad -\mathbb{E}_{p_D(x^1,\cdots,x^N)}\sum_{i=1}^N
            D_{KL}\left[ q(z^s|x^1,\cdots,x^N) \| r^i(z^s|x^i)
        \right] \notag \\
    & \quad\quad -\sum_{i=1}^N
            \mathbb{E}_{p_D(x^i)}D_{KL}\left[q(z^i|x^i) \| p(z^i)
        \right] + \sum_{i=1}^NH(X^i) \,.
\end{align}
Clearly, there are many terms also present in the ELBO~(Eq.~\ref{eq:ELBO}). 
We aim to learn a generative model by maximizing the log-likelihood of the joint multilingual distribution while enforcing disagreement by maximizing the regularization in Eq.~\ref{eq:disent_bound}.
Thus, we combine these objectives using a trade-off weight \(\lambda\):
\begin{align}\label{eq:opt_obj}
    & \mathbb{E}_{q(z^1,\cdots,z^N,z^s, x^1,\cdots,x^N)}\left[
            \log \frac{p(x^1,\cdots,x^N,z^1,\cdots,z^N,z^s)}{q(z^1,\cdots, z^N, z^s | x^1, \cdots, x^N)}
        \right] \notag \\
     & \quad\quad + \lambda \sum^N_{i=1}\left[ I(X^i;\overline{X^i};Z^s) - I(Z^i;Z^s) \right]  \notag \\
     & \geq (1+\lambda)\mathbb{E}_{p_D(x^1,\cdots,x^N)}
            \sum_{i=1}^N \mathbb{E}_{q(z^i|x^i)q(z^s|x^1,\cdots,x^N)}[\log p(x^i|z^i,z^s)] \notag  \\
     & \quad\quad - (1+\lambda) \mathbb{E}_{p_D(x^i)} \sum_{i=1}^N D_{KL}\left[ q(z^i|x^i)||p(z^i) \right] \\
     & \quad\quad - \mathbb{E}_{p_D(x^i)} D_{KL}\left[q(z^s | x^1, \cdots, x^N) \| p(z^s) \right] \notag  \\
     & \quad\quad - \lambda \cdot \mathbb{E}_{p_\mathcal{D}(x^1,\cdots,x^N)}[\sum_{i=1}^N D_{KL} [q(z^s|x^1,\cdots,x^N)||r^i(z^s_i|x^i)]] \notag \,.
\end{align}
The right-hand side~(RHS) of Eq.~\ref{eq:opt_obj} represents the objective for multilingual program translation within a disentangled generative framework.

\subsection{The overall framework of VIM-PT}
The VIM-PT is architecture-free, which can theoretically and practically be adapted to arbitrary sequence-to-sequence architecture, which involves an encoder and a decoder. Based on the experimental performance, we finally set up a unified encoder for all the languages with weight-sharing and an independent decoder for each programming language.
As illustrated in Figure~\ref{framework}, given a multi-parallel sample $(t^1,\cdots,t^N)$ from the dataset $p_\mathcal{D} (t^1,\cdots,t^N)$, where each sample including a group of codes of the same function implemented by the N different programming languages. The flag tokens are initialized by the unused tokens in the tokenizer. To refer to the programming language, we concatenate the flag tokens, and source code as the input sequence as follows:
\begin{equation}
    [[CLS], x^n_1, x^n_2, \cdots, x^n_k, t^n_1, t^n_2, \cdots, t^n_c, [SEP]] \text{,} n \in [1, N] \text{,}
\end{equation}
where k is the length of the flag token, and c is the length of the source code.
Then the sequence of the instance is encoded by the weight-sharing encoders to the initial representations. Then the flag representations are inputted into the variational interaction.

After variational interaction, we can reconstruct the flag representations of the other programming languages for each source code of the sample. Then the target flag representations are concatenated with the source code representations and inputted into the target decoder to generate the target code. 
The regularized training approach learns the model from the objective which consists of the supervised loss function as well as the regularization term described in Eq.~\ref{eq:opt_obj}.

\subsection{Learning from multi-parallel samples}
The disentangle module involves three approximate posteriors $q_{\phi_i}$, $r_{\phi_i}$, and $p_{\phi_i}$, and a language-shared approximate posterior $q_{\phi_s}$. Each approximate posterior is parameterized by a standard Gaussian prior $z\sim \mathcal{N}(0, I)$.

The flag representations of instances can be disentangled by the corresponding language-specific projectors $(q_{\phi_1}, \cdots, q_{\phi_N})$ and language-shared projectors $q_{\phi_s}$, respectively. 
In addition, the shift-shared representations are inferred by corresponding shift-shared projectors $(r_{\phi_1}, \cdots, r_{\phi_N})$.
Then the language-specific representations and the language-shared representation are inputted into $(p_{\phi_1}, \cdots, p_{\phi_N})$ to reconstruct the representations of the multi-lingual instances. 

The language-specific representations, the language-shared representation, and the shift-shared representations are denoted by $(\textbf{z}^1,\cdots,\textbf{z}^N)$, $\textbf{z}^s$, and $(\textbf{z}^s_1, \cdots, \textbf{z}^s_N)$, respectively. 
Instructed by Equation~\ref{eq:opt_obj}, the objective of the training of multi-parallel samples is formalized as:

\begin{align}\label{train}
\mathcal{L} & = ( (1+\lambda) \cdot [ - \sum_{i=1}^N \mathbb{E}_{q(\textbf{z}^i|\textbf{x}^i)q(\textbf{z}^s|\textbf{x}^1,\cdots,\textbf{x}^N)}[\underbrace{\log p(\textbf{x}^i|\textbf{z}^i,\textbf{z}^s)}_\text{Reconstruction Loss} \notag \\
& \quad \quad  + \underbrace{\log\mathit{p}_D(\mathbf{t}^i|\mathbf{\hat{x}}^i, \mathbf{t}^s)] ])}_\text{Translation Cross Entropy}  \\
 &\quad + (1+\lambda)\cdot \sum_{i=1}^N D_{KL}[q(\textbf{z}^i|\textbf{x}^i)||p(\textbf{z}^i)] \notag \\
 &\quad + D_{KL}[q(\textbf{z}^s|\textbf{x}^1,\cdots, \textbf{x}^N)||p(\textbf{z}^s)] \notag \\
 &\quad +\lambda \cdot \sum_{i=1}^N D_{KL}[q(\textbf{z}^s|\textbf{x}^1,\cdots,\textbf{x}^N)||r^i(\textbf{z}^s_i|\textbf{x}^i)] \notag  \,,
\end{align}

The first term of the Eq.~\ref{train} includes two parts, where the reconstruction loss (Mean Square Error) and translation cross-entropy loss are to reconstruct the representations of flag tokens and source code tokens, respectively. The last term is to minimize the KL divergence between the shift-shared and language-shared representations, so the language-shared representation can be replaced approximately by the shift-shared representation when some instances are missing in the partially missing samples. 

\subsection{Learning from partially missing samples}
For partially missing samples, we design a training strategy to exploit the data that is not multi-parallel.
The training strategy is described in Algorithm~\ref{algo}.

At first, given a partially missing sample from the dataset, there may be some missing instances in one sample. To obtain the pseudo instances, we disentangle the language-specific representations and the language-shared representations to generate the draft translation of the missing instances.
We sample a set of pseudo instances $(\widetilde{t}^1,\cdots,\widetilde{t}^N)$ from the training dataset randomly to fill the missing instances of the corresponding language in the original sample.
And the flag representations can be separated as $(\widetilde{\mathbf{x}}^1,\cdots,\widetilde{\mathbf{x}}^N)$.
Similar to the learning of multi-parallel samples, the language-specific representations, and the shift-shared representations are disentangled as 
$(\widetilde{\textbf{z}}^1,\cdots,\widetilde{\textbf{z}}^N)$, and $(\widetilde{\textbf{z}}^s_1, \cdots, \widetilde{\textbf{z}}^s_N)$. However, as some instances of the sample are missing, the language-shared representation can not be inferred.

\begin{algorithm}[h]
\small
\KwIn{
Dataset $\mathcal{D}$ from N programming languages, $opt$ is the optimizer
}
\KwOut{Model($\theta, \phi$)}

\For{$epoch < max\_epoch$}
{   Draw a partially missing sample from the dataset\\
    Construct the pseudo sample $(\widetilde{t}^1,\cdots,\widetilde{t}^N)$ by filling the missing instance from training set randomly \\
    Encode the sample by encoders as $(\widetilde{\mathbf{x}}^1,\cdots,\widetilde{\mathbf{x}}^N)$ \\
    {\color{blue} Enforce disentanglement on pseudo instances} to obtain language-specific representation $(\widetilde{\textbf{z}}^1,\cdots,\widetilde{\textbf{z}}^N)$ and shift-shared representation $(\widetilde{\textbf{z}}^s_1, \cdots, \widetilde{\textbf{z}}^s_N)$ by $(q_{\phi_1}, \cdots, q_{\phi_N})$ and $(r_{\phi_1}, \cdots, r_{\phi_N})$, respectively \\
    Compute $\sum_{i=1}^N D_{KL}[q_{\phi_i}(\widetilde{\textbf{z}}^i|\widetilde{\mathbf{x}}^i)||p(\widetilde{\textbf{z}}^i)]$ \\
    \For{each $\widetilde{\textbf{z}}^s_i$ obtained by $r_{\phi_i}$}
    {  Reconstruct the sample $(\mathbf{x}^1,\cdots, \mathbf{x}^N)$ with $(\widetilde{\textbf{z}}^1,\cdots,\widetilde{\textbf{z}}^N)$ and $\widetilde{\textbf{z}}^s_i$ by $(p_{\phi_1}, \cdots, p_{\phi_N})$\\ 
    {\color{blue} Enforce disentanglement on multi-parallel instances} to obtain language-specific representation $(\textbf{z}^1,\cdots,\textbf{z}^N)$ and language-shared representation $\textbf{z}^s$ by $(q_{\phi_1}, \cdots, q_{\phi_N})$ and $q_{\phi_s}$ respectively \\
    Compute $\sum_{i=1}^N D_{KL}[q_{\phi_i}(\textbf{z}^i|\mathbf{x}^i)||p(\textbf{z}^i)]$, $D_{KL}[q_{\phi_s}(\textbf{z}^s|\textbf{x}^1,\cdots, \textbf{x}^N)||p(\textbf{z}^s)]$ and $\sum_{i=1}^N D_{KL}[q_{\phi_s}(\textbf{z}^s|\mathbf{x}^1,\cdots, \mathbf{x}^N)||r_{\phi_i}(\textbf{z}^s|\mathbf{x}^i)]$ \\
    Reconstruct the instance representation $(\mathbf{\hat{x}}^1,\cdots, \mathbf{\hat{x}}^N)$ with $(\textbf{z}^1,\cdots,\textbf{z}^N)$ and $\textbf{z}^s$ by $(p_{\phi_1}, \cdots, p_{\phi_N})$  \\ 
    Generate the target programs with ground truth by decoders and compute the first term of Eq.~\ref{train}.  
    }
    $\theta, \phi \gets opt(\theta, \phi, \nabla_{\theta, \phi})$ \\
}
\caption{Pseudocode of training of partially missing samples.\label{algo}}
\end{algorithm}

To overcome this issue, the method generates the draft translation of the missing instances with the pseudo instances. Each language-specific representation of the pseudo instance and the language-shared representation of ground truth are inputted into $(p_{\phi_1}, \cdots, p_{\phi_N})$ to reconstruct the representations of the multi-lingual instances. 
As the language-shared representation of ground truth is not unique, for each language-shared representation of $(\widetilde{\textbf{z}}^s_1, \cdots, \widetilde{\textbf{z}}^s_N)$, a multi-parallel sample $(\mathbf{\hat{x}}^1,\cdots, \mathbf{\hat{x}}^N$ reconstructed by language-shared information and language-specific information can be obtained. 
Then the learning of multi-parallel samples can be performed. Especially, only the target programs with ground truth are reconstructed.

\begin{table*}[ht]
\small
    \centering
    \caption{\label{data1}
    Sample counts for each language in CoST. Py denotes Python, and JS denotes JavaScript.
    }
    \begin{tabular}{cccccccccc}
    \toprule
       \textbf{Granularity} & \textbf{Dataset} & \textbf{Java} & \textbf{C\#} & \textbf{C++} & \textbf{C} & \textbf{Py} & \textbf{PHP} & \textbf{JS} & \textbf{Samples} \\
    \midrule
       \multirow{3}{*}{Program}  & train & 1,442 & 1,382 & 1,442 & 183 & 1,343 & 435 & 904 & 1,507 \\
                                 & val  & 49 & 49 & 49 & 49 & 49 & 49 & 49 & 49 \\
                                 & test  & 69 &  69 &  69 &  69 &  69 &  69 &  69 &  69 \\
    \midrule
       \multirow{3}{*}{Snippet}  & train & 13,979 & 47,722 & 27,101 & 4,517 & 34,582 & 12,752 & 28,376 & 49,962 \\
                                 & val  & 272 & 1,524 & 544 & 813 & 948 & 1,402 & 1,185 & 1,524 \\
                                 & test  & 411 & 2,301 & 823 & 1,238 & 1,423 & 2,138 & 1,778 & 2,308 \\
    \bottomrule
    \end{tabular}
    
\end{table*}

\begin{figure*}[ht]
\begin{minipage}{0.28\linewidth}
        \includegraphics[width=2in, height=2in]{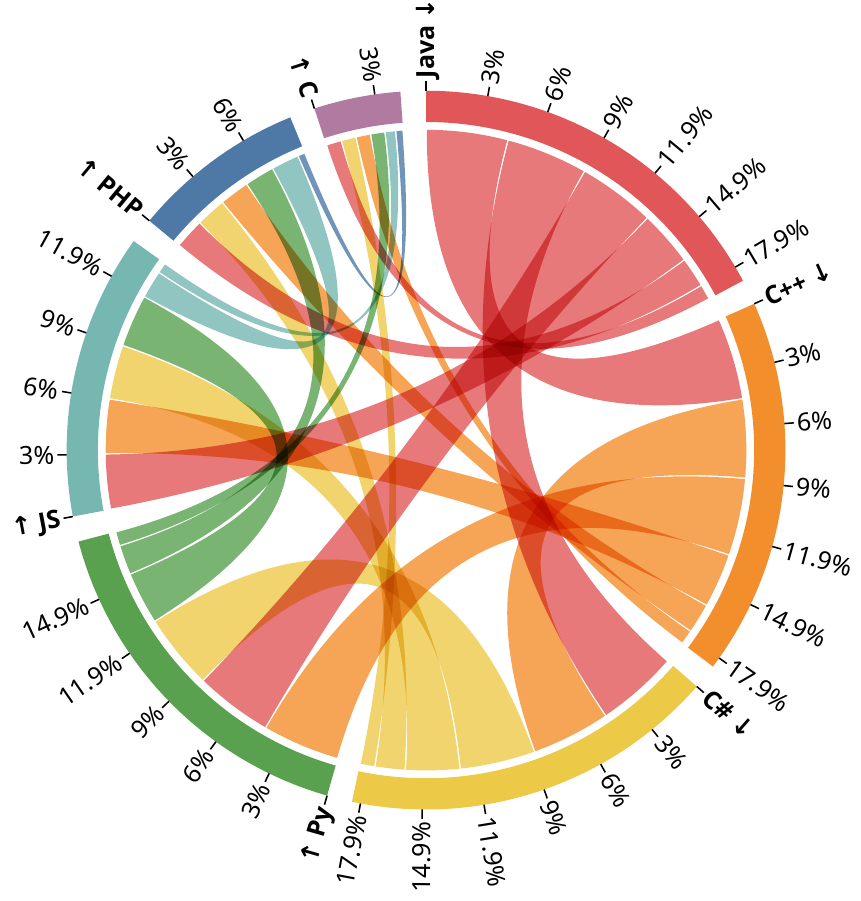}
        \captionof{figure}{\label{fig:cost_pro_pairs}
        	Program Pairs on CoST.
        }
\end{minipage}
\begin{minipage}{0.28\linewidth}
        \includegraphics[width=2in, height=2in]{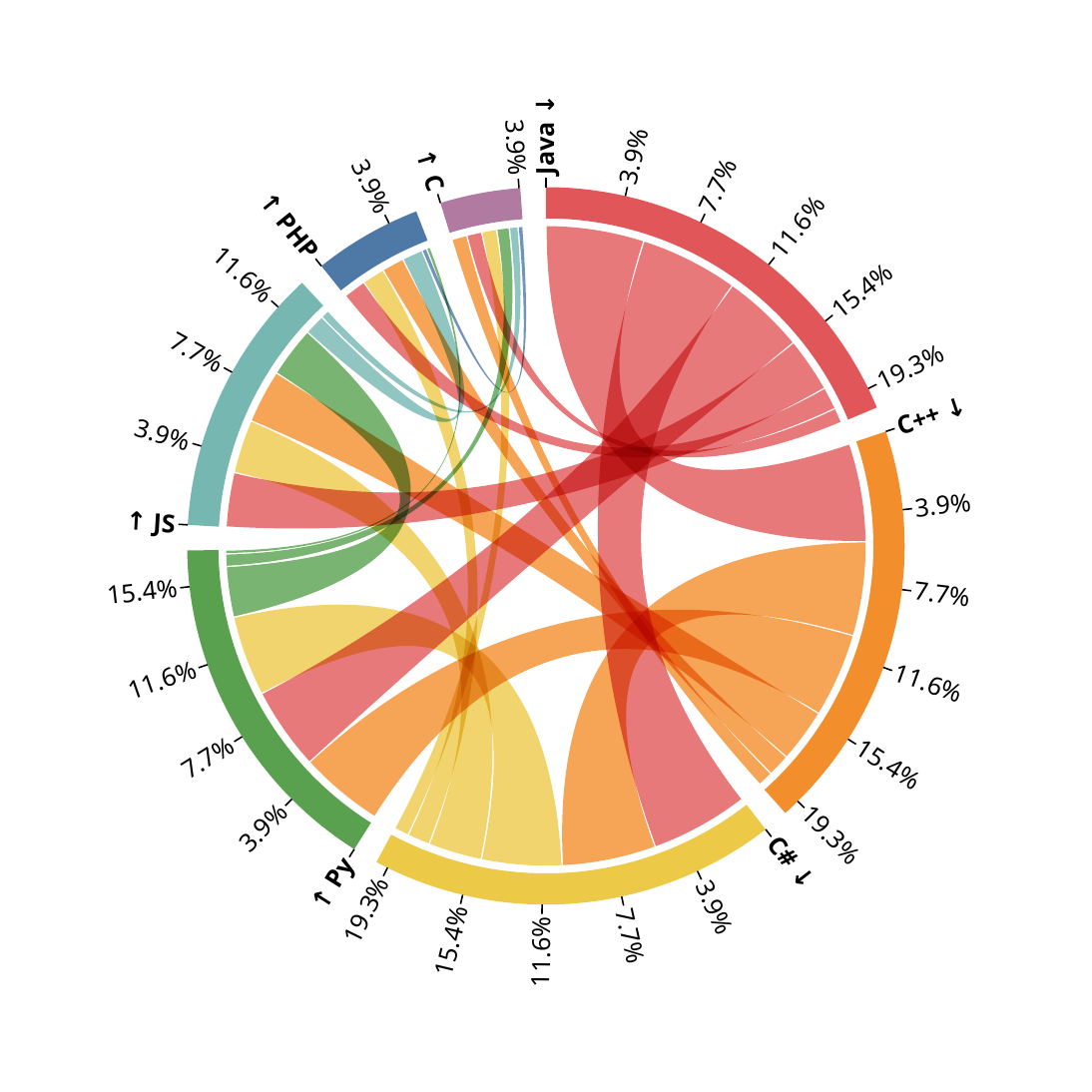}
        \captionof{figure}{\label{fig:cost_sni_pairs}
        	Snippet Pairs on CoST.
        }
\end{minipage}
\begin{minipage}{0.37\linewidth}
    \captionof{table}{\label{data2}
        Pairwise statistics of the CoST. The upper triangle~(in normal font) shows the number of parallel snippets, while the lower triangle~(in bold font) shows the number of parallel programs. 
        JS represents JavaScript and Py represents Python.
    }
    \footnotesize
	\begin{tabular}{lccccccc}
    \toprule
       \textbf{Lang} & 
            \cellcolor[HTML]{B07AA1}\textbf{\textcolor{white}{C}}    & 
            \cellcolor[HTML]{EDC948}\textbf{\textcolor{white}{C\#}}  & 
            \cellcolor[HTML]{F28E2B}\textbf{\textcolor{white}{C++}}  & 
            \cellcolor[HTML]{E15759}\textbf{\textcolor{white}{Java}} & 
            \cellcolor[HTML]{76B7B2}\textbf{\textcolor{white}{JS}}   & 
            \cellcolor[HTML]{4E79A7}\textbf{\textcolor{white}{PHP}}  & 
            \cellcolor[HTML]{59A14F}\textbf{\textcolor{white}{Py}}   \\
    
       \cellcolor[HTML]{B07AA1}\textbf{\textcolor{white}{C}}    &  -  & 2123 & 2188 & 2135 & 1232 & 700 & 1779 \\
       \cellcolor[HTML]{EDC948}\textbf{\textcolor{white}{C\#}}  & \textbf{273} & -   & 13326 & 13905 & 7601 & 3192 & 11404 \\
       \cellcolor[HTML]{F28E2B}\textbf{\textcolor{white}{C++}}  & \textbf{267} & \textbf{1442} & -     & 13929 & 7596 & 3165 & 11930 \\
       \cellcolor[HTML]{E15759}\textbf{\textcolor{white}{Java}} & \textbf{281} & \textbf{1495} & \textbf{1497}  &  -  & 7729 & 3194 & 11713 \\  
        \cellcolor[HTML]{76B7B2}\textbf{\textcolor{white}{JS}}  & \textbf{196} & \textbf{994} & \textbf{996} & \textbf{1009} &   -    & 2917 & 7165 \\
       \cellcolor[HTML]{4E79A7}\textbf{\textcolor{white}{PHP}}  & \textbf{135} & \textbf{552} & \textbf{548} &  \textbf{552}  & \textbf{512}  &  -  & 545 \\
       \cellcolor[HTML]{59A14F}\textbf{\textcolor{white}{Py}}   & \textbf{263} & \textbf{1383} & \textbf{1419} & \textbf{1417} & \textbf{962} & \textbf{545}  &  -   \\
    \bottomrule
    \end{tabular}
\end{minipage}
\end{figure*}

\section{EXPERIMENTS}
\subsection{Dataset Description}
We conduct experiments on the \textit{CoST} dataset~\cite{must}, which is a large and comprehensive dataset to evaluate the performance of program translation approaches.
The dataset consists of both snippet-level and program-level parallel data from 7 programming languages~(i.e., C, C\#, C++, Java, JavaScript, PHP, and Python) and up to 42 programming language pairs, which was collected from the GeeksForGeeks website\footnote{\url{https://www.geeksforgeeks.org/}}. 
The platform ensures its contributors stick to a template in terms of the comments used in their programs and the code corresponding to those comments. 
The dataset provides a good number of multilingual instances of code that can be effectively used for this task. 
The detailed statistics of the dataset are shown in Table~\ref{data1}, while the train, validation, and test sets are split the same as \textit{CoST}~\cite{must}. 
For each language, there are some instances of corresponding samples that are missing, especially in C and PHP.

Fig.~\ref{fig:cost_pro_pairs} and Fig.~\ref{fig:cost_sni_pairs} have illustrated the imbalanced proportion of the pairwise data in the dataset. The statistics of the pairwise data are shown in the Table~\ref{data2} in detail.

\begin{table*}[t]
\footnotesize
    \centering
    \caption{Performance evaluation in terms of BLEU-4.}
    \resizebox{\textwidth}{!}{
    \begin{tabular}{ll|ccccccc|ccccccc}
        \toprule
        & & \multicolumn{7}{c|}{Snippet Level} & \multicolumn{7}{c}{Program Level} \\
        \midrule
        Lang & Method & C & C\# & C++ & Java & JS & PHP & Py & C & C\# & C++ & Java & JS & PHP & Py \\
        \midrule
        \multirow{7}{*}{C} & Na\"ive Copy & - & 68.88 & 85.58 & 69.17 & 54.85&37.48 &37.84 & - & 68.74 & \cellcolor{highlight2}85.02 & 69.34 &53.90 &38.1 & 36.09 \\
        & DOBF & - &39.38 &40.17 &41.57 &33.54 &34.51 &17.93  & - & 27.64 & 20.90 & 27.15 & 20.77 & 24.15 & 25.87 \\
        & CodeBERT & - &51.92 &60.84 &51.70 &40.57 &34.49 &31.49  & - & 34.41 & 33.64 & 33.75 & 29.90 & 29.32 & 27.34 \\
        & MuST-PT & - &\cellcolor{highlight2}80.68 & \cellcolor{highlight2}88.58 &\cellcolor{highlight2}79.24 &\cellcolor{highlight2}80.35 & \cellcolor{highlight2}82.94 &\cellcolor{highlight2}66.49  & - & \cellcolor{highlight2}78.39 & \cellcolor{highlight3}84.92 & \cellcolor{highlight2}76.84 & \cellcolor{highlight3}66.13 & \cellcolor{highlight2}70.62 & \cellcolor{highlight3}55.71 \\
        & M-PT(\textit{w/o} VI) & - & \cellcolor{highlight3}76.82 & \cellcolor{highlight3}84.65 & \cellcolor{highlight3}70.57 & \cellcolor{highlight3}73.86 & \cellcolor{highlight3}72.83 & \cellcolor{highlight3}65.94 & - & \cellcolor{highlight3}74.76 & 76.22 & \cellcolor{highlight3}72.28 & \cellcolor{highlight2}70.91 & \cellcolor{highlight3}69.57 & \cellcolor{highlight2}59.62 \\ 
        & VIM-PT & - & \cellcolor{highlight}81.21 & \cellcolor{highlight}89.80 & \cellcolor{highlight}79.85 & \cellcolor{highlight}80.98 & \cellcolor{highlight}85.86 & \cellcolor{highlight}69.47 & - & \cellcolor{highlight}81.19 & \cellcolor{highlight}85.64 & \cellcolor{highlight}78.83 & \cellcolor{highlight}73.81 & \cellcolor{highlight}73.82 & \cellcolor{highlight}65.48 \\
        \midrule
        \multirow{7}{*}{C\#} & Na\"ive Copy  & 68.91 & - &67.29 &78.13 &58.90 &35.01 &36.49 & 68.74 & - & 67.51 & 78.69 & 57.61 & 35.55 & 34.62 \\
        & DOBF & 38.33 & - &38.94 &47.84 &28.70 &49.32 &25.14 & 26.38 &- & 27.50 & 31.63 & 23.62 & 34.90 & 22.94 \\
        & CodeBERT& 52.65 & - &79.28 &\cellcolor{highlight3}83.90 &76.99 &6 &64.72   & 34.93 & -& 65.74 & 80.11 & 53.72 & 45.67 & 47.14 \\
        & MuST-PT & \cellcolor{highlight2}81.12 & - & \cellcolor{highlight2}85.34 & \cellcolor{highlight}85.80 & \cellcolor{highlight2}82.74 & \cellcolor{highlight2}81.64 & \cellcolor{highlight3}71.11 & \cellcolor{highlight2}78.78 &- & \cellcolor{highlight2}84.72 & \cellcolor{highlight}87.76 & \cellcolor{highlight3}70.00 & \cellcolor{highlight2}70.66 & \cellcolor{highlight2}62.03 \\
        & M-PT(\textit{w/o} VI) & \cellcolor{highlight3}79.41 & - & \cellcolor{highlight3}81.24 & 83.75 & \cellcolor{highlight3}78.64 & \cellcolor{highlight3}75.83 & \cellcolor{highlight2}72.16 & \cellcolor{highlight3}69.53 & - & \cellcolor{highlight3}74.87 & \cellcolor{highlight3}80.80 & \cellcolor{highlight2}71.63 & \cellcolor{highlight3}68.38 & \cellcolor{highlight3}61.09 \\
        & VIM-PT & \cellcolor{highlight}82.02 & - & \cellcolor{highlight}85.92 & \cellcolor{highlight2}83.99 & \cellcolor{highlight}83.15 & \cellcolor{highlight}85.27& \cellcolor{highlight}73.42 & \cellcolor{highlight}79.36 & - & \cellcolor{highlight}84.82  & \cellcolor{highlight2}87.63 & \cellcolor{highlight}81.41 & \cellcolor{highlight}74.26 & \cellcolor{highlight}65.78 \\
        \midrule
        \multirow{7}{*}{C++} & Na\"ive Copy & 85.66 & 67.33 & -  & 67.32&55.44 &37.68 &36.92 & \cellcolor{highlight}85.02 & 67.51 & -& 67.38 & 54.07 & 38.47 & 34.89 \\
        & DOBF & 43.32 & 42.25 & - &42.03 &40.01 &49.29 &25.77 & 31.84 & 37.43 &- & 48.70 & 34.05 & 15.67 & 23.73 \\
        & CodeBERT &63.24 &77.21 & - &78.39 &75.10 &70.75 &68.92  & 45.57 & 64.15 & -& 56.47 & 48.65 & 40.59 & 56.73 \\
        & MuST-PT & \cellcolor{highlight}87.55 & \cellcolor{highlight3}82.98 & - & \cellcolor{highlight3}80.27 & \cellcolor{highlight2}81.01 & \cellcolor{highlight2}83.29 & \cellcolor{highlight2}71.20 & \cellcolor{highlight2}84.20 & \cellcolor{highlight2}81.15 &- & \cellcolor{highlight2}79.15 & \cellcolor{highlight2}68.85 & \cellcolor{highlight2}71.18 & \cellcolor{highlight2}64.10 \\
        & M-PT(\textit{w/o} VI) & \cellcolor{highlight3}83.40 & \cellcolor{highlight2}83.54 & - & \cellcolor{highlight2}81.67 & \cellcolor{highlight3}78.91 & \cellcolor{highlight3}77.71 & \cellcolor{highlight3}69.92 & 71.75 & \cellcolor{highlight3}74.43 & - & \cellcolor{highlight3}72.57 & \cellcolor{highlight2}71.19 & \cellcolor{highlight3}69.94 & \cellcolor{highlight3}61.27 \\
        & VIM-PT & \cellcolor{highlight2}86.57 & \cellcolor{highlight}86.19 & - & \cellcolor{highlight}84.69 & \cellcolor{highlight}82.35 & \cellcolor{highlight}83.56 & \cellcolor{highlight}73.21 & \cellcolor{highlight3}77.11 & \cellcolor{highlight}83.27 & - & \cellcolor{highlight}79.92 & \cellcolor{highlight}75.63 & \cellcolor{highlight}73.89 & \cellcolor{highlight}66.07 \\
        \midrule
        \multirow{7}{*}{Java} & Na\"ive Copy &69.14 &78.03 &67.25 & - &57.33 &33.82 &35.50 & 69.40 & 78.77& 67.48 & - & 55.99 & 33.66 & 33.60 \\
        & DOBF & 39.21&44.26 & 38.80 & - & 40.23&48.87 &24.83 & 32.23 & 65.02 & 22.01 &- & 55.78 & 35.75 & 24.90 \\
        & CodeBERT  & 54.98& 86.02 & 79.14 & - &78.54 &70.21 &66.41 & 46.85 & \cellcolor{highlight3}80.88 & 69.88 & - & 55.15 & 47.66 & 48.56 \\
        & MuST-PT & \cellcolor{highlight2}81.16 & \cellcolor{highlight3}90.13  & \cellcolor{highlight2}85.23 & - & \cellcolor{highlight2}81.87 & \cellcolor{highlight}80.39 & \cellcolor{highlight3}70.06 & \cellcolor{highlight2}78.71 & \cellcolor{highlight2}89.93 & \cellcolor{highlight2}84.28 &- & \cellcolor{highlight3}69.53 & \cellcolor{highlight2}69.83 & \cellcolor{highlight3}61.12 \\
        & M-PT(\textit{w/o} VI) & \cellcolor{highlight3}78.54 & \cellcolor{highlight2}90.36 & \cellcolor{highlight3}81.07 & - & \cellcolor{highlight3}80.55 & \cellcolor{highlight3}72.31 & \cellcolor{highlight2}74.03 & \cellcolor{highlight3}70.25 & 78.66 & \cellcolor{highlight3}75.54 & - & \cellcolor{highlight2}70.80 & \cellcolor{highlight3}69.20 & \cellcolor{highlight2}61.59 \\
        & VIM-PT & \cellcolor{highlight}81.96 & \cellcolor{highlight}93.12 & \cellcolor{highlight}85.79 & - & \cellcolor{highlight}83.29 & \cellcolor{highlight2}80.25 & \cellcolor{highlight}78.29 & \cellcolor{highlight}82.09 & \cellcolor{highlight}90.43 & \cellcolor{highlight}84.79 & - & \cellcolor{highlight}75.89 & \cellcolor{highlight}74.25 & \cellcolor{highlight}66.04 \\
        \midrule
        \multirow{7}{*}{JS} & Na\"ive Copy &54.58 &58.61 &55.29  &56.61 & - &30.44 & 41.58 & 53.00 & 56.70 &53.29 & 54.44&-& 31.53 & 39.77 \\
        & DOBF & 33.16& 41.50& 39.91 & 44.16 & - & 46.93& 24.05   &22.13 &38.06 &20.69 &37.78 &- &26.03 &21.21 \\
        & CodeBERT  & 40.93& 75.38& 73.83 & 74.58 & - & 63.85& 60.99  &32.88 & 58.43 & 50.42 &60.13 & -& 46.14 & 44.34 \\
        & MuST-PT  & \cellcolor{highlight2}78.54 & \cellcolor{highlight3}78.91 & \cellcolor{highlight}78.95 & \cellcolor{highlight3}78.03 & - & \cellcolor{highlight2}78.69 & \cellcolor{highlight3}66.47  & \cellcolor{highlight2}70.20 & \cellcolor{highlight2}73.32 & \cellcolor{highlight2}73.01 & \cellcolor{highlight2}73.39 &- & \cellcolor{highlight2}76.44 & \cellcolor{highlight2}63.88 \\
        & M-PT(\textit{w/o} VI) & \cellcolor{highlight3}75.90 & \cellcolor{highlight2}80.44 & \cellcolor{highlight3}77.40 & \cellcolor{highlight2}79.23 & - & \cellcolor{highlight3}78.56 & \cellcolor{highlight2}66.53 & \cellcolor{highlight3}65.99 & \cellcolor{highlight3}71.93 & \cellcolor{highlight3}72.57 & \cellcolor{highlight3}70.08 & - & \cellcolor{highlight3}64.68 & \cellcolor{highlight3}56.61 \\
        & VIM-PT & \cellcolor{highlight}79.96 & \cellcolor{highlight}81.76 & \cellcolor{highlight2}78.09 & \cellcolor{highlight}79.27 & - & \cellcolor{highlight}79.51 & \cellcolor{highlight}67.02 & \cellcolor{highlight}70.64 & \cellcolor{highlight}79.07 & \cellcolor{highlight}75.05 & \cellcolor{highlight}74.07 & - & \cellcolor{highlight}77.57 & \cellcolor{highlight}64.72 \\
        \midrule
        \multirow{7}{*}{PHP} & Na\"ive Copy & 37.46& 34.99& 37.64 & 33.85 & 30.66 & - & 23.67  &38.10 & 35.55 & 38.47 & 33.61 & 32.01 & -& 23.04 \\
        & DOBF &25.78 &40.88 & 38.30 & 42.98 & 38.11& - & 25.64& 17.62 & 31.95 & 30.18 & 25.88 & 25.89 &- & 20.80 \\
        & CodeBERT  &30.06 &65.67 & 67.68 & 64.02 & 62.06& - & 57.01& 30.82 & 47.14 & 43.04 & 45.83 & 43.45 &- & 39.42 \\
        & MuST-PT & \cellcolor{highlight2}76.67 &\cellcolor{highlight2}77.96 & \cellcolor{highlight2}79.41 & \cellcolor{highlight2}76.42 & \cellcolor{highlight}77.64 & - & \cellcolor{highlight2}69.34 & \cellcolor{highlight3}67.88 & \cellcolor{highlight3}70.34 & \cellcolor{highlight3}70.04 & \cellcolor{highlight3}67.30 & \cellcolor{highlight2}73.54 &- & \cellcolor{highlight2}63.97 \\
        & M-PT(\textit{w/o} VI) & \cellcolor{highlight3}73.68 & \cellcolor{highlight3}75.04 & \cellcolor{highlight3}76.02 & \cellcolor{highlight3}68.10 & \cellcolor{highlight3}68.20 & - & \cellcolor{highlight3}61.25 & \cellcolor{highlight2}68.33 & \cellcolor{highlight2}72.24 & \cellcolor{highlight2}73.42 & \cellcolor{highlight2}70.49 & \cellcolor{highlight3}69.12 & - & \cellcolor{highlight3}61.08 \\
        & VIM-PT & \cellcolor{highlight}78.30 & \cellcolor{highlight}78.79 & \cellcolor{highlight}81.37 & \cellcolor{highlight}77.02 & \cellcolor{highlight2}76.76 & - & \cellcolor{highlight}69.65 & \cellcolor{highlight}70.58 & \cellcolor{highlight}73.82 & \cellcolor{highlight}80.30 & \cellcolor{highlight}73.64 & \cellcolor{highlight}76.76 & - & \cellcolor{highlight}70.44 \\
        \midrule
        \multirow{7}{*}{Py} & Na\"ive Copy & 37.77& 36.42& 36.90 & 35.24 & 41.53 & 23.59 & - & 35.74 & 34.31 &34.62 & 33.00&  39.79&  22.85 & - \\
        & DOBF & 28.77& 34.07& 36.23 & 33.48 & 30.71 & 45.68 & -& 15.47 & 29.22 & 24.99 & 35.64 & 27.31 & 28.21 &- \\
        & CodeBERT  &35.82 & 61.50& 71.06 & 65.99 & 62.34 & 63.73 & -& 53.17 & 49.16 & 57.63 & 52.93 & 52.30 & 45.69 &- \\
        & MuST-PT & \cellcolor{highlight2}70.64 & \cellcolor{highlight2}72.35 & \cellcolor{highlight2}75.37 & \cellcolor{highlight2}70.89 & \cellcolor{highlight2}70.46  & \cellcolor{highlight2}75.49 & - & \cellcolor{highlight3}58.70 & \cellcolor{highlight3}63.23 & \cellcolor{highlight3}66.16 & \cellcolor{highlight3}64.57 & \cellcolor{highlight2}66.47 & \cellcolor{highlight2}70.90 &- \\
        & M-PT(\textit{w/o} VI) & \cellcolor{highlight3}63.72 & \cellcolor{highlight3}67.88 & \cellcolor{highlight3}72.39 & \cellcolor{highlight3}67.54 & \cellcolor{highlight3}69.43 & \cellcolor{highlight3}66.78 & - & \cellcolor{highlight2}63.72 & \cellcolor{highlight2}67.88 & \cellcolor{highlight2}68.39 & \cellcolor{highlight2}67.54 & \cellcolor{highlight3}59.43 & \cellcolor{highlight3}66.78 & - \\
        & VIM-PT & \cellcolor{highlight}71.11 & \cellcolor{highlight}77.68 & \cellcolor{highlight}75.85 & \cellcolor{highlight}73.45 & \cellcolor{highlight}72.74 & \cellcolor{highlight}76.05 & - & \cellcolor{highlight}64.65 & \cellcolor{highlight}68.53 & \cellcolor{highlight}69.59 & \cellcolor{highlight}67.64 & \cellcolor{highlight}68.29 & \cellcolor{highlight}71.00 & - \\
        \bottomrule
    \end{tabular}
    }
    \label{SOTA}
\end{table*}

\begin{table*}[t]
\small
    \centering
    \caption{The average BLEU-4 score of the translation of different approaches \textit{from} or \textit{to} the languages.}
    \resizebox{\textwidth}{!}{
    \begin{tabular}{l|cccccccc|cccccccc}
        \toprule
        Method & \multicolumn{8}{c|}{Snippet Level} & \multicolumn{8}{c}{Program Level} \\
        \midrule
        \textit{from} & C & C\# & C++ & Java & JS & PHP & Py & Avg & C & C\# & C++ & Java & JS & PHP & Py & Avg \\
        \midrule
        Na\"ive & 58.97 & 57.46 & 58.39 & 56.85 & 49.52 & 33.05 & 35.24 & 49.93 & 58.53 & 57.12 & 57.89 & 56.48 & 48.12 & 33.46 & 33.39 & 49.28 \\
        DOBF & 34.52 & 38.05 & 40.45 & 39.37 & 38.29 & 35.28 & 34.82 & 37.25 & 24.41 & 27.83 & 31.90 & 39.28 & 27.65 & 25.39 & 26.81 & 29.04 \\
        CodeBERT & 45.17 & 71.03 & 72.27 & 72.55 & 64.93 & 57.75 & 60.07 & 63.40 & 31.39 & 54.55 & 52.03 & 58.16 & 48.72 & 41.62 & 51.81 & 48.33 \\
        MuST-PT & \cellcolor{highlight2}79.71 & \cellcolor{highlight2}81.29 & \cellcolor{highlight2}81.05 & \cellcolor{highlight2}81.47 & \cellcolor{highlight2}76.60 & \cellcolor{highlight2}76.24 & \cellcolor{highlight2}72.53 & \cellcolor{highlight2}78.41 & \cellcolor{highlight2}72.10 & \cellcolor{highlight2}75.66 & \cellcolor{highlight2}74.77 & \cellcolor{highlight2}75.57 & \cellcolor{highlight2}71.71 & \cellcolor{highlight3}68.85 & \cellcolor{highlight3}65.01 & \cellcolor{highlight2}71.95 \\
        M-PT(\textit{w/o} VI) & \cellcolor{highlight3}74.11 & \cellcolor{highlight3}78.50 & \cellcolor{highlight3}79.19 & \cellcolor{highlight3}79.48 & \cellcolor{highlight3}76.34 & \cellcolor{highlight3}70.38 & \cellcolor{highlight3}67.96 & \cellcolor{highlight3}75.14 & \cellcolor{highlight3}70.56 & \cellcolor{highlight3}71.05 & \cellcolor{highlight3}70.19 & \cellcolor{highlight3}71.01 & \cellcolor{highlight3}66.98 & \cellcolor{highlight2}69.11 & \cellcolor{highlight2}65.62 & \cellcolor{highlight3}69.22 \\
        VIM-PT & \cellcolor{highlight}81.12 & \cellcolor{highlight}82.30 & \cellcolor{highlight}82.76 & \cellcolor{highlight}83.78 & \cellcolor{highlight}77.60 & \cellcolor{highlight}76.98 & \cellcolor{highlight}74.48 & \cellcolor{highlight}79.86 & \cellcolor{highlight}76.46 & \cellcolor{highlight}78.88 & \cellcolor{highlight}75.98 & \cellcolor{highlight}78.92 & \cellcolor{highlight}73.52 & \cellcolor{highlight}74.26 & \cellcolor{highlight}68.28 & \cellcolor{highlight}75.19 \\
        \textit{Improvement}(\%) & 9.46 & 4.84 & 4.51 & 5.41 & 1.65 & 9.38 & 9.59 & 6.41 & 8.36 & 11.02 & 8.25 & 11.14 & 9.76 & 7.45 & 4.05 & 8.58 \\
        \midrule
        \midrule
        \textit{to} & C & C\# & C++ & Java & JS & PHP & Py & Avg & C & C\# & C++ & Java & JS & PHP & Py & Avg \\
        \midrule
        Na\"ive & 58.92 & 57.38 & 58.33 & 56.72 & 49.79 & 33.00 & 35.33 & 49.92 & 58.33 & 56.93 & 57.73 & 56.08 & 48.90 & 33.36 & 33.67 & 49.29 \\
        DOBF  & 34.76 & 40.39 & 38.73 & 42.01 & 35.22 & 45.77 & 23.89 & 37.25 & 24.28 & 38.22 & 24.38 & 34.46 & 31.24 & 27.45 & 23.24 & 29.04 \\
        CodeBERT  & 46.28 & 69.62 & 71.97 & 69.76 & 65.93 & 61.94 & 58.26 & 63.39 & 40.70 & 55.70 & 53.39 & 54.87 & 47.20 & 42.51 & 43.92 & 48.33 \\
        MuST-PT  & \cellcolor{highlight2}79.28 & \cellcolor{highlight2}80.50 & \cellcolor{highlight2}82.15 & \cellcolor{highlight2}78.44 & \cellcolor{highlight2}79.01 & \cellcolor{highlight2}80.41 & \cellcolor{highlight2}69.11 & \cellcolor{highlight2}78.41 & \cellcolor{highlight2}73.08 & \cellcolor{highlight2}76.06 & \cellcolor{highlight2}77.19 & \cellcolor{highlight2}74.84 & \cellcolor{highlight2}69.09 & \cellcolor{highlight2}71.61 & \cellcolor{highlight2}61.80 & \cellcolor{highlight2}71.95 \\
        M-PT(\textit{w/o} VI)  & \cellcolor{highlight3}75.78 & \cellcolor{highlight3}79.01 & \cellcolor{highlight3}78.80 & \cellcolor{highlight3}75.14 &\cellcolor{highlight3} 74.93 & \cellcolor{highlight3}74.00 & \cellcolor{highlight3}68.31 & \cellcolor{highlight3}75.14 & \cellcolor{highlight3}68.26 &\cellcolor{highlight3}73.32 & \cellcolor{highlight3}73.50 & \cellcolor{highlight3}72.29 & \cellcolor{highlight3}68.85 & \cellcolor{highlight3}68.09 & \cellcolor{highlight3}60.21 & \cellcolor{highlight3}69.22 \\
        VIM-PT  & \cellcolor{highlight}79.99 & \cellcolor{highlight}83.13 & \cellcolor{highlight}82.80 & \cellcolor{highlight}79.71 & \cellcolor{highlight}79.88 & \cellcolor{highlight}81.75 & \cellcolor{highlight}71.84 & \cellcolor{highlight}79.87 &\cellcolor{highlight} 74.07 & \cellcolor{highlight}79.39 & \cellcolor{highlight}80.03 & \cellcolor{highlight}76.96 & \cellcolor{highlight}75.30 & \cellcolor{highlight}74.13 & \cellcolor{highlight}66.42 & \cellcolor{highlight}75.19 \\
        \textit{Improvement}(\%) & 5.56 & 5.20 & 5.09 & 6.08 & 6.60 & 10.47 & 5.18 & 6.31 & 8.51 & 8.28 & 8.88 & 6.45 & 9.37 & 8.87 & 10.32 & 8.67 \\
        \bottomrule
    \end{tabular}
    }
    \label{avgscore}
\end{table*}

\begin{table}[t]
    \centering
    \caption{Comparison with LLMs on program-level, including CodeT5, Deepseek-Coder(1.3B), and GPT-3.5.}
    \small
    \begin{tabular}{l|cccccc}
    \toprule
    Model & Py-Java & C\#-Java & C\#-Py & C++-C\# & C\#-C++ \\
    \midrule
    CodeT5 & 66.40 & 82.41 & 62.80 & 72.47 & 78.39 \\
    Deepseek-Coder & 54.13 & 52.97 & 59.00 & 55.07 & 53.03 \\
    GPT-3.5 & 65.60 & 85.64 & 62.93 & 79.01 & 81.31 \\
    VIM-PT & 67.64 & 87.63 & 65.78 & 83.27 & 84.82 \\
    \bottomrule
    \end{tabular}
    \label{llm}
\end{table}

\subsection{Experimental Settings}
We choose the following state-of-the-art multi-lingual and pairwise program translation methods as baselines. All the encoder-only pre-trained models are fine-tuned on the task with the same decoder as our methods.
\begin{itemize}[leftmargin=*]
    \item \textbf{Na\"ive Copy}: A direct copy from the source program to the translation output, which denotes how similar the source language and the target language are.
    \item \textbf{DOBF} \cite{DOBF}: A pre-training model for programming languages, which leverages the structural aspect of programming languages and recovers the original version of obfuscated source code.
    \item \textbf{CodeBERT} \cite{codebert}: A bimodal pre-training model, trained with a hybrid objective function that incorporates the pre-training task of standard masked language modeling and replaced token detection in the pre-training stage. 
    \item \textbf{MuST-PT} \cite{must}: The state-of-the-art multilingual program translation technique that utilizes multilingual languages and exhibits strong generalizability enhances translation performance, particularly for low-resource languages.
    \item \textbf{M-PT(\textit{w/o} VI)}: A straightforward approach of multilingual program translation. where each programming language shares an encoder and has an independent decoder. This approach is a blank controller of VIM-PT without variational interaction.
\end{itemize}

To compare with these baselines, we follow the best hyper-parameters suggested in their studies. For hyper-parameters in our method, the numbers of transformer layers of the encoder and decoder are set as 12 and 6, respectively. The model dimension and attention heads in transformer layers are set as 768 and 12. 
The hyper-parameters $\alpha$, $k$ are determined based on the performance of the validation set, which is set as 1e-3, 16 in the experiment. Then, the training set and the validation set are mixed up to train the model. Following MuST-PT\cite{must}, the snippet-level training set is utilized to enhance the program-level translation, and some data of C++ is used to supplement C. Following the baselines~\cite{Transcoder, must}, the comments in the source code are kept in the dataset, which can increase the number of anchor points across languages, and the parameters of CodeBERT~\cite{codebert} are utilized to initialize the encoder to accelerate the training process.
The AdamW~\cite{AdamW} optimizer is used to update model parameters with the initial learning rate 1e-4. The linear weight decay is used for scheduling the learning rate.
The process is repeated for $3$ times and the average performance on the test set is reported.
All the experiments are conducted with the NVIDIA Tesla A100 with 128GB RAM on the Ubuntu system.

\begin{figure*}[t]
    \centering
    \includegraphics[width=\linewidth]{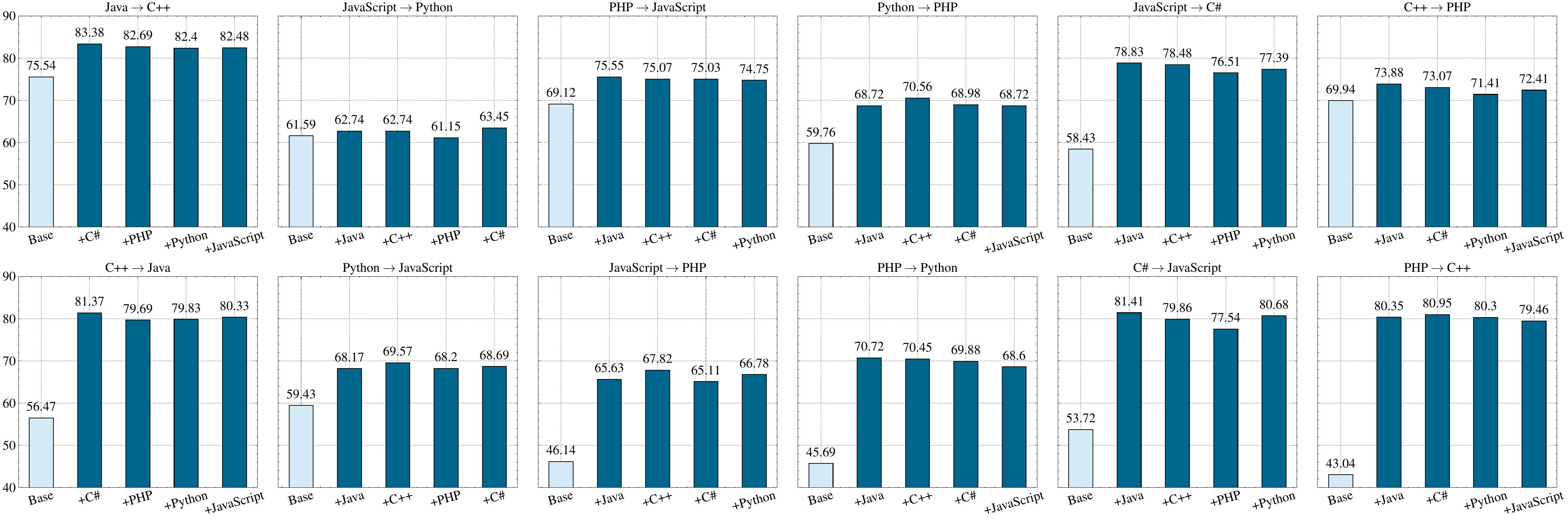}
    \caption{The auxiliary benefit of different languages in program translation.}
    \label{auxiliary}
\end{figure*}
\label{-0.3cm}

\subsection{RQ1: What is the Performance of VIM-PT Comparing With Baseline Approaches?}
In the experiment, we use the BLEU~\cite{Bleu} score as the evaluation metric to evaluate the $n$-gram overlap between the translated code and the ground-truth target code, which is the most widely used metric in program translation~\cite{GraphCodeBERT,must,DOBF,CodeXGLUE}.
A higher BLEU score indicates better evaluation performance, which varies from 0 to 100 as a percentage.
Table~\ref{SOTA} shows the experimental results. 
For each pair of translations, the three best performing approaches are highlighted, while dark gray marks an approach with the highest retrieval performance, and gray and light gray correspond to the second- and third-best approaches.
Compared with the pairwise approach like CodeBERT with a similar backbone, the multilingual approaches including MuST-PT, VIM-PT(w/o VT), and VIM-PT perform significantly outperformance. We can attribute this result to (1) expansion of the representation scope, as the different language corpora face various application scenarios and have different functional preferences, (2) better latent space, which can be jointly constructed by different languages to represent the language-sharing semantics, and (3) the auxiliary benefit to low-resource language with rich-resource language, which is last but most important factor. In practical scenarios, where different language corpora are unevenly distributed, building a low-resource programming language to latent space mapping can be benefited from learning among other rich-resource programming languages, which is more effective and robust than building a pairwise mapping using only parallel data.

It can be observed that VIM-PT performs performance gains widely over the state-of-the-art approaches on most of the translation pairs, especially at the program level. VIM-PT performs better with the improvement of 4.57\% and 1.87\% over the state-of-the-art approach in multi-lingual program translation~\cite{must} at the program level and snippet level, respectively.
In particular, VIM-PT performs better than all the pairwise approaches and multi-lingual approaches on all the translations when Python is the source language or target language.
Table~\ref{avgscore} further calculates the average score of the translation of different approaches \textit{from} or \textit{to} any programming languages based on Table~\ref{SOTA}.
As shown in Table~\ref{avgscore}, VIM-PT has performed the state-of-the-art at both snippet and program level in all configurations of the average score \textit{from} or \textit{to} any programming languages. Additionally, the p-values between VIM-PT and each baseline are less than 0.05, confirming that the observed differences are statistically significant. We notice the performance of MusT-PT is lower than M-PT(\textit{w/o} VI) on PHP and Python, which indicates that using the weight-sharing decoder is not conducive to the modeling of low-resource languages with few samples, which may be overwhelmed by the rich-resource languages in the joint training.

Moreover, we further evaluate the performance of large language models on the program translation. As shown in Table~\ref{llm}, CodeT5~\cite{codet5} and VIM-PT are trained by supervised fine-tuning, while Deepseek-Coder(1.3B)~\cite{guo2024deepseek} and GPT-3.5~\cite{ModelsOp75:online} are evaluated by inference. Limited by space, we only present some of the experimental results. As shown in Table~\ref{llm}, GPT-3.5 performs well across language pairs with direct inference, whereas the performance of Deepseek-Coder(1.3B) is comparatively lower, which can be attributed to differences in model size. But VIM-PT achieves comparable results to GPT-3.5 despite having significantly fewer parameters.

\subsection{RQ2: What is the impact of Variational Interaction for the VIM-PT?}
To evaluate the effectiveness of the variational interaction, we set a blank controller termed M-PT(\textit{w/o} VI), which ablates variational interaction on VIM-PT.
As shown in Table~\ref{SOTA} and Table~\ref{avgscore}, the inclusion of Variational Interaction (VI) in VIM-PT results in an improvement in the average BLEU-4 score at both snippet Level and program Level. Across most programming language pairs, VIM-PT achieves higher performance than M-PT(\textit{w/o} VI), indicating that variational interaction indeed enhances translation quality. 

In summary, the incorporation of VI effectively enhances the performance of the translation model, resulting in more accurate and fluent translations.
The \textit{Improvement} in Table~\ref{avgscore} indicate the average improvement of the variational interaction across all languages and levels, with an average increase from 4.05\% to 11.14\%, and from 1.65\% to 10.47\% at the program level and snippet level, respectively.
The variational interaction leads to notable improvements especially in low-resource languages like C and PHP, indicating that variational interaction plays a crucial role in enhancing the performance of the low-resource languages in multilingual translation.

\subsection{RQ3: Which rich-resource languages benefit the low-resource languages?}
Due to the diversity between different programming languages, we explore the auxiliary benefits of different programming languages as rich-resource to the translation between low-resource programming languages.
As shown in Fig.~\ref{auxiliary}, with auxiliary of any other programming languages, the performance of each direction of program translation has obtained a significant improvement. It is because when importing the third programming language, more data are utilized in the learning and the language-shared latent space can be constructed better.
The base indicates the pair-wised approach based on CodeBERT and the others are the translation with the help of the other languages based on the framework of VIM-PT.
The auxiliary benefits of different programming languages may depend on the following possible factors.

\subsubsection{The imbalance of the data of different auxiliary languages}
The amount of data of auxiliary language has a significant influence on the auxiliary benefit of program translation.
As mentioned above, more data are utilized in the learning and the language-shared latent space can be constructed better, and further influence the decoding of the language-shared features.
As illustrated in Fig.~\ref{fig:cost_pro_pairs}, compared to PHP the number of pairs from Java, C++, and C\# is larger. Therefore, when Java, C++, and C\#  are selected as auxiliary languages, the benefits for the translation are more significant than PHP as the auxiliary language generally.

\subsubsection{Compiled and interpreted programming languages}
Programming languages can be divided into compiled and interpreted types, depending on how programs are executed. Compiled languages require that all source code be translated into machine code when writing a program, which is then executed by a computer. Common compiled languages include C, C++, and Java. Interpreted languages translate source code into machine code line by line or block by block when executing a program and execute it. Common interpreted languages include Python, JavaScript, and PHP.
As shown in Fig.~\ref{auxiliary}, in the compiled-compiled translation from C++ to Java, using the compiled language C\# as auxiliary can achieve higher performance than the other interpreted languages, such as PHP, Javascript, and Python.

\subsubsection{Syntax Similarity}
Although each programming language has a unique formal syntax, the syntax similarity between different languages is various.
For instance, C\# is similar to C++ in syntax, while quite different from Python.
As shown in Fig.~\ref{auxiliary}, in the translation from PHP to C++, with the help of C\#, the improvement of the performance is more significant than Python as an auxiliary.

In summary, it is observed that any language that serves as an auxiliary generally has a supporting role, which demonstrates the effectiveness of VIM-PT in addressing the distribution shift of semantics across multiple languages. 
Yet many possible factors influence the benefit of the auxiliary language, thus how to select the rich-resource languages as auxiliary languages is still a question worth exploring in the multi-lingual program translation.

\subsection{RQ4: What is the deployment complexity of the joint model and pairwise approaches?}
\begin{figure}[t]
    \centering
    \includegraphics[width=0.8\linewidth]{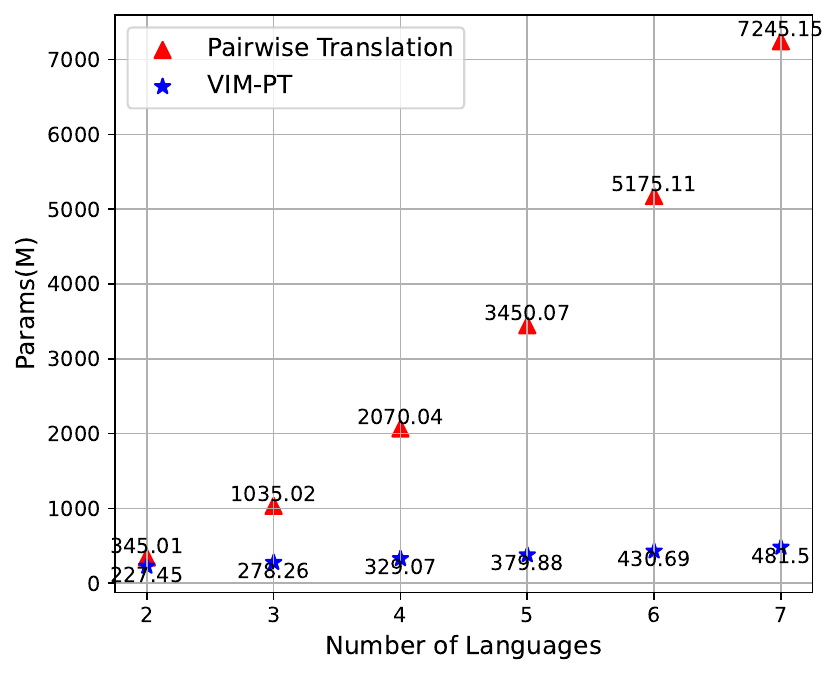}
    \caption{The comparison of space overhead between pairwise approach and VIM-PT in terms of parameters.}
    \label{size}
\end{figure}

To compare the deployment complexity between the pairwise approach and VIM-PT, we record the space overhead of the translation model. As shown in Fig.~\ref{size}, the pairwise approach and VIM-PT are both constructed with CodeBERT as encoder and the 6-layer Transformer as Decoder. With the number of languages in translation increasing, the number of parameters of the pairwise approach grows exponentially, and VIM-PT grows linearly.

It is because the encoder-decoder models needed by the pairwise approach are coupling with the number of translation directions, but the decoders needed by VIM-PT are coupling with the number of languages. 
In the bilingual program translation, the pairwise approach should construct two models for the two directions in translation, and the number of parameters of the pairwise approach and VIM-PT is similar.
When the multilingual translation gradually increases to 7 languages, the pairwise approach should construct 42 encoder-decoder models for the 42 directions in translation. In this case, VIM-PT only needs one weight-sharing encoder and 7 decoders, where the number of parameters is less than 1/15 of the pairwise approaches. It demonstrates the space efficiency of multilingual program translation.

\section{Threat Analysis}
\label{threat}
Our results are interpreted with two threats to validity in mind.
\begin{itemize}[leftmargin=*]
    \item The internal threat to validity lies in the implementation of compared techniques. To reduce it, we directly reuse the implementation of the compared techniques from their reproducible packages and the weights of pre-trained models, if they are available and executable.
    Otherwise, we reimplement the techniques strictly following the papers on existing mature libraries.
    \item The external threat to validity lies in the dataset used in the experiment. To mitigate the external threat, the widely used dataset, which includes 7 programming languages and up to 42 pairs and two levels, is used to evaluate the effectiveness of the method. The artifact contains the dataset and the code is publicly available in the supplementary material.
\end{itemize}

\section{Conclusion}

In this paper, we argue that jointly learning a unified model to translate code across multiple programming languages is superior to separately learning from bilingual parallel data. We propose Variational Interaction for Multilingual Program Translation (VIM-PT), a disentanglement-based generative approach that jointly trains a unified model for multilingual program translation across multiple languages.
VIM-PT disentangles code into language-shared and language-specific features using variational inference and interaction information with a novel lower bound. With the variational interaction, VIM-PT achieves significant improvement in multilingual program translation, mines and leverages the capability of non-parallel data, addresses the distribution shift of program semantics across languages, and serves as a unified model, reducing deployment complexity. 
In the future, more effective selection strategies in filling the partially missing samples can be explored beyond the random selection used in our method.

\section{Data Availability}
Our replication package (including code, model, etc.) is publicly available at \url{https://github.com/duyali2000/VIM-PT}.

\section{Acknowledgments}
This research was supported by NSFC~(62076121, 61921006), Major Program~(JD) of Hubei Province~(2023BAA024), and Postgraduate Research \& Practice Innovation Program of Jiangsu Province (KYCX24\_0301). The authors would like to thank Hao-Yuan He for his helpful feedback on drafts of the paper.


\appendix
\section{Appendices}
\subsection{$I(Z^i; Z^s) = -I(X^i; Z^i, Z^s) + I(X^i; Z^i) + I(X^i; Z^s)$}
\label{IZIZS}
With the interaction information between three random variables:
\begin{equation}
    I(X;Y;Z) = I(X;Z) - I(X;Z|Y) = I(Y;Z) - I(Y;Z|X) \,,
\end{equation}
we can obtain the mutual information between $Z^i$ and $Z^s$:
\begin{equation}
    I(Z^i, Z^s) = I(Z^i; X^i) - I(Z^i;X^i|Z^s) + I(Z^i;Z^s|X^i)\,.
\end{equation}
Due to the structural assumption on $q$, the $q(z^i|x^i) = q(z^i|x^i,z^s)$ holds, so the last term in the above function is eliminated: 
\begin{equation}
\begin{aligned}
    I(Z^i;Z^s|X^i) & = H(Z^i|X^i) - H(Z^i|X^i,Z^s) \\
    & = H(Z^i|X^i) - H(Z^i|X^i) = 0\,,
\end{aligned}
\end{equation}
which yields
\begin{equation}
\begin{aligned}
     I(Z^i, Z^s) & = I(X^i; Z^i) - I(X^i;Z^i|Z^s)  \\
     & = I(X^i; Z^i) + I(X^i; Z^s) - I(X^i; Z^i, Z^s)\,.
\end{aligned}
\end{equation}

\bibliographystyle{ACM-Reference-Format}
\bibliography{sample-base}

\end{document}